# Quantum Coulomb Blockade in Orbital Resolved Phosphorus Triple-Donor Molecule

Soumya Chakraborty,*[1] Pooja Sudha,*[1] Hemant Arora,[1] Daniel Moraru,[2] & Arup Samanta[1,3,a)]

[1]Quantum / Nano-science and Technology Lab, Department of Physics, Indian Institute of Technology Roorkee, Roorkee 247667, Uttarakhand, India

[2]Research Institute of Electronics, Shizuoka University, 3-5-1 Johoku, Chuo-ku, Hamamatsu 432-8011, Japan

[3]Centre of Nanotechnology, Indian Institute of Technology Roorkee, Roorkee 247667, Uttarakhand, India

[a)]arup.samanta@ph.iitr.ac.in, and *Equal contribution

Multi-donor architecture in silicon offers a promising direction towards scalable solid-state qubits and quantum technologies operating at practical conditions. However, the overlap of multiple donor wave-functions develops a complex internal electronic configuration with several discrete energy levels. Probing these discrete-correlated states is essential for understanding inter-donor coupling and exchange interactions towards their practical implementations in quantum-technologies. We have experimentally demonstrated quantum Coulomb blockade mediated systematic filling of several electrons into orbital-resolved molecular states within multi-phosphorous-donor molecules accompanied by a correlated decrement in charging energies for higher hybridized orbitals due to expanded Bohr radii and electron delocalization. Corresponding, first-principle density functional theory calculations offer microscopic insight into the orbital configurations, while the rate equation simulations of quantum Coulomb blockade faithfully reproduce the experimental stability diagrams. This comprehensive characterization advances and discusses the role of donor-molecules in silicon in scalable building blocks for quantum technologies operable at elevated temperatures.

## I. INTRODUCTION

Quantum technologies, including quantum computers and advanced quantum-electronic devices, fundamentally rely on precise control of single electrons within quantum-confined systems. Scalable donor-based silicon (Si) architectures[1-11] are the cutting-edge advanced technology due to their atomic precision, enhanced stability, and compatibility with existing semiconductor technologies.[12] Seamless operation of such sophisticated devices at elevated temperatures, against thermal fluctuations, requires sustained single-electron effects, but most reports are limited to temperatures of a few tens of Kelvin.[12-16] Although the leverage of confinement effects[17,18] has facilitated the enhanced barrier heights along with high charging-energy ($E_{ch}$) and to harness the high-temperature single-electron tunneling exploiting even via isolated donors,[19-21] effectiveness and reproducibility of this approach is heavily impaired by the inherent challenges of nano-fabrication and precise dopant placement. A more effective strategy is to create a deep-level quantum dot (QD) by coupling a few closely-spaced donor-atoms, where the ground-state is found at energy far lower than the isolated-donor systems, resulting in enhanced electron charging energy and ionization energy ($E_I$), necessary for sustained Coulomb blockade (CB) at higher temperatures.[22-24]

However, in the closely-spaced multiple-donor molecule, donor's wave functions will superimpose to each other, resulting in the splitting of individual-donor energy levels to a many-fold quantized energy spectrum. The ability to probe such discrete electronic states, often resolved as multiple conductance peaks in the transport characteristics of such devices, is critical for understanding the complex interplay between confinement potential and inter-donor couplings[25] and exchange interactions.[26-28] Moreover, understanding the systematic one-by-one electron transport and filling within such a system with evolving external bias voltages is essential for coherent single-electron manipulation towards qubit operations. Although spin-blockade and quantum transport through multi-donor molecule have been reported,[22,24] a systematic investigation of electron filling and the effect of each molecular energy level on the charging and ionization energies is highly essential to develop the controlled quantum devices, as reported for the shell filling of electrons in a GaAs QD system.[29]

Here, we experimentally demonstrate the systematic one-by-one molecular filling of six electrons, following the Quantum Coulomb Blockade (QCB), in the lowest-lying hybridized energy levels of the phosphorus (P) triple-donor molecule formed within a Si nano-channel field-effect transistor (nano-FET). Moreover, the observed QCB synchronizes precisely with the systematic evolution of available discrete excited states and the respective separation of energy levels ($\Delta E$) that actively influences the transport characteristics. Furthermore, a systematic decrement of $E_{ch}$ has been observed, corresponding to higher energy levels with increasing Bohr radius ($r_B$) within each molecule. The presented systematic investigation of the origin and dynamics of such donor-molecular states supported by ab-initio and quantum Coulomb blockade simulations provides a rich understanding of multi-donor molecule-based high-temperature quantum devices.

## II. METHODS

### A. Device fabrication

The device is fabricated on the silicon-on-insulator (SOI) platform comprised of a buried oxide (BOX) layer of 150 nm thickness and an initial silicon device layer of 55 nm thickness. Upon patterning the constricted nano-wire channel with length and width of approximately 180 nm and 15 nm, respectively, by an electron-beam lithography technique, a 30 nm × 15 nm opening was created, 90 nm away from the source and drain reservoir, which was further doped with P-donor concentration of $N_D \approx 10^{19}$ cm$^{-3}$.Through the standard and sacrificial oxidation processes during the fabrication steps, the device layer is reduced to approximately 5 nm. Such heavily-doped nano-slit contains approximately 22 donors, with the average inter-donor spacing of ~ 1.5 nm, suggesting strong interaction among the donors to facilitate multi-donor cluster formation. All fabrication processes are conducted in a clean-room environment and follow CMOS-compatible methodologies. After the channel patterning, the top oxide layer of thickness approximately 10 nm is grown by thermal dry oxidation at 800°C for 15 minutes, as a final gate oxide. Aluminum (Al) was used for the gate, source, and drain contacts.

## B. Electrical measurements

Electrical measurements are conducted utilizing an Agilent 4156C precision semiconductor parameter analyzer coupled to a variable-temperature probe station. All current-voltage (*I*-*V*) measurements are performed in high vacuum. During all the measurements, the source and substrate remain grounded, while bias is applied to the drain electrode as $V_{DS}$. Top-gate is used to control the charge-states of the donors in the channel region, by the gate voltage $V_G$.

## C. Theoretical calculations

We performed Density Functional Theory (DFT) calculations using Quantum-ATK software.[30] First, we constructed a silicon nanowire with a diameter of approximately 1.6 nm and a length of 5.4 nm. To minimize the influence of surrounding atoms, we introduced a 1 nm vacuum region around the nanowire. Three P-donors were substituted into the Si matrix, as illustrated in Fig. 4a, representing a plausible experimental scenario for the 3-donor molecules. The nanowire dimensions are smaller than those in typical experimental devices due to computational resource limitations.

# III. RESULTS and DISCUSSION

## A. Donor potential distribution within a thin Si nano-channel

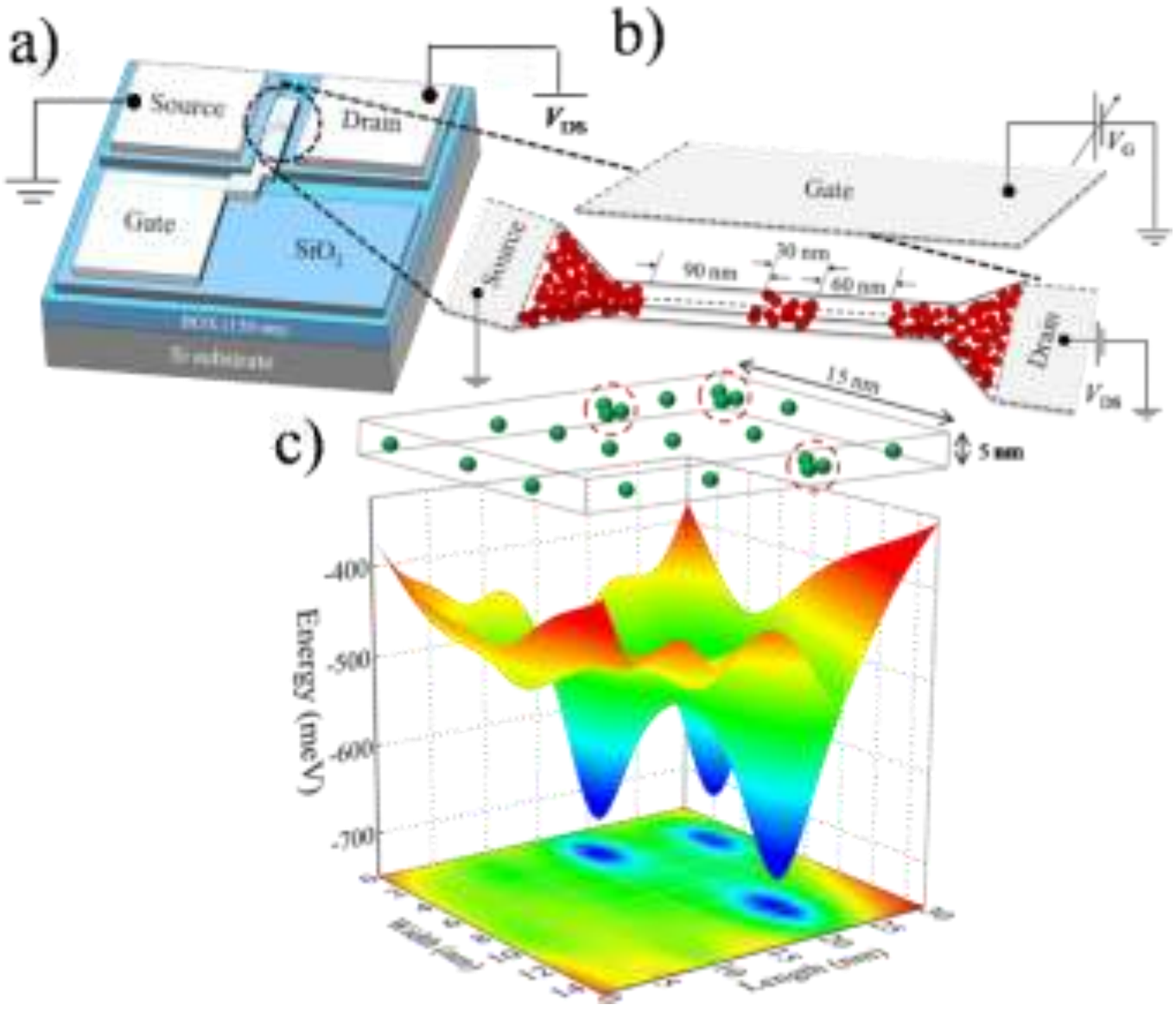

**Figure 1 Schematic Si nano-FET device configuration. a**, Representation of the fabricated device as a silicon-on-insulator field-effect transistor (SOI-FET). **b**, Corresponding donor distribution within the channel region along with the top-gate design. **c**, An instance of donor distribution profile in the slit region of the nano-channel by employing a Poisson distribution approach. The corresponding probable potential landscape is simulated by approximating each isolated P-donor with a truncated Coulomb potential of the form $ktanh(\lambda r)/r$, within the spherical shell effective mass approximation. The values for $k = 1.67$ and $1/\lambda = 1.7$ nm are variationally set so that the isolated P-donor ground state energy converges to the bulk value of ~45 meV with an effective $r_B$ of 2.4 nm.

In this regard, we demonstrate QCB mediated systematic filling of several single electrons into orbital resolved triple-donor molecule in a Si nano-FET.

The channel dimension of the device is ≈180 nm length, ≈15 nm width, ≈5 nm thickness as schematically depicted in Figs. 1a-b. A precisely defined 30 nm × 15 nm region, 90 nm from the source, is selectively doped with phosphorus (P) donor concentration $N_D \approx 10^{19}$ cm$^{-3}$, resulting in an average of ≈ 22 randomly distributed donors (Figs. 1b-c). Such doping-configuration with inter-donor spacing of ≈1.5 nm, significantly smaller than twice the Bohr radius ($2r_B \approx 5$ nm) for P-donor in bulk Si, would favor strong inter-donor coupling, enhancing the formation of donor clusters with artificial molecular-like states.[22,25]

An instance of donor distribution along-with the corresponding donor-induced potential landscape, exclusive of confinement effects and considering random donor-distribution, of the slit region are simulated in Fig. 1c.

Therein, we can observe an overall conduction-band lowering, around 400 meV, in the slit region due to the interaction among all the 22 closely-spaced donors. This is additionally superimposed by three isolated potential dips of ≈ 300 meV each. These isolated potential-dips are due to further clustering of three donor-atoms, marked by red-dotted circles in Fig. 1c. Probabilistic calculation of cluster-formation for 22 Poisson-distributed donors,[31] with $r_B \approx 2.4$ nm, within the mentioned device geometry would also suggest the formation of three triple-donor molecules, as found in the randomly distributed donor potential simulation of Fig. 1c.

## B. Experimental observation of orbital-resolved QCB

To characterize the device, we measure the corresponding stability diagram, depicting the evolution of the differential-conductance $dI_{DS}/dV_{DS}$ as a function of the bias voltage - gate voltage ($V_{DS}$ -$V_G$) plane at $T = 5.5$ K (Fig. 2). The stability diagram reveals a well-organized and recurring pattern of six groups of single-electron peaks (Group-1 to Group-6), each containing three distinct peaks. Low gate-bias single-electron peaks are accompanied by cross-biasing effects, where Coulomb diamonds become tilted and non-closed due to cross-talk between gate and source/drain.[32] Such complicated and repetitive Coulomb oscillations cannot be correlated with transport via isolated donor atoms.[33,34] In general, observation of a large number of periodic Coulomb-oscillations can also be observed in geometrical QD induced by roughness, which is not in the present scenario as the gap between the successive Coulomb-peaks does not follow the QCB of a single QD and secondly, the formation of such a small roughness-induced-QD in the present devices is highly negligible.[35] The observation of several groups of Coulomb-oscillations generally arises from series interactive QDs or parallel QDs. In case of series interaction, charge transport via interactive QDs can modify the number of Coulomb peaks along with the random variation in charging energies, which is not the present scenario.[35-37] Another possible option is the transport through parallel non-interactive different QDs. [33,34] Considering the large periodicity of the Coulomb-oscillations, QDs must be very small in sizes. Formation of such small-sized geometrical QDs by a few numbers only is highly unlikely in the present device fabrication process. Furthermore, considering the selectively doped channel region, formation of such QDs by interaction of

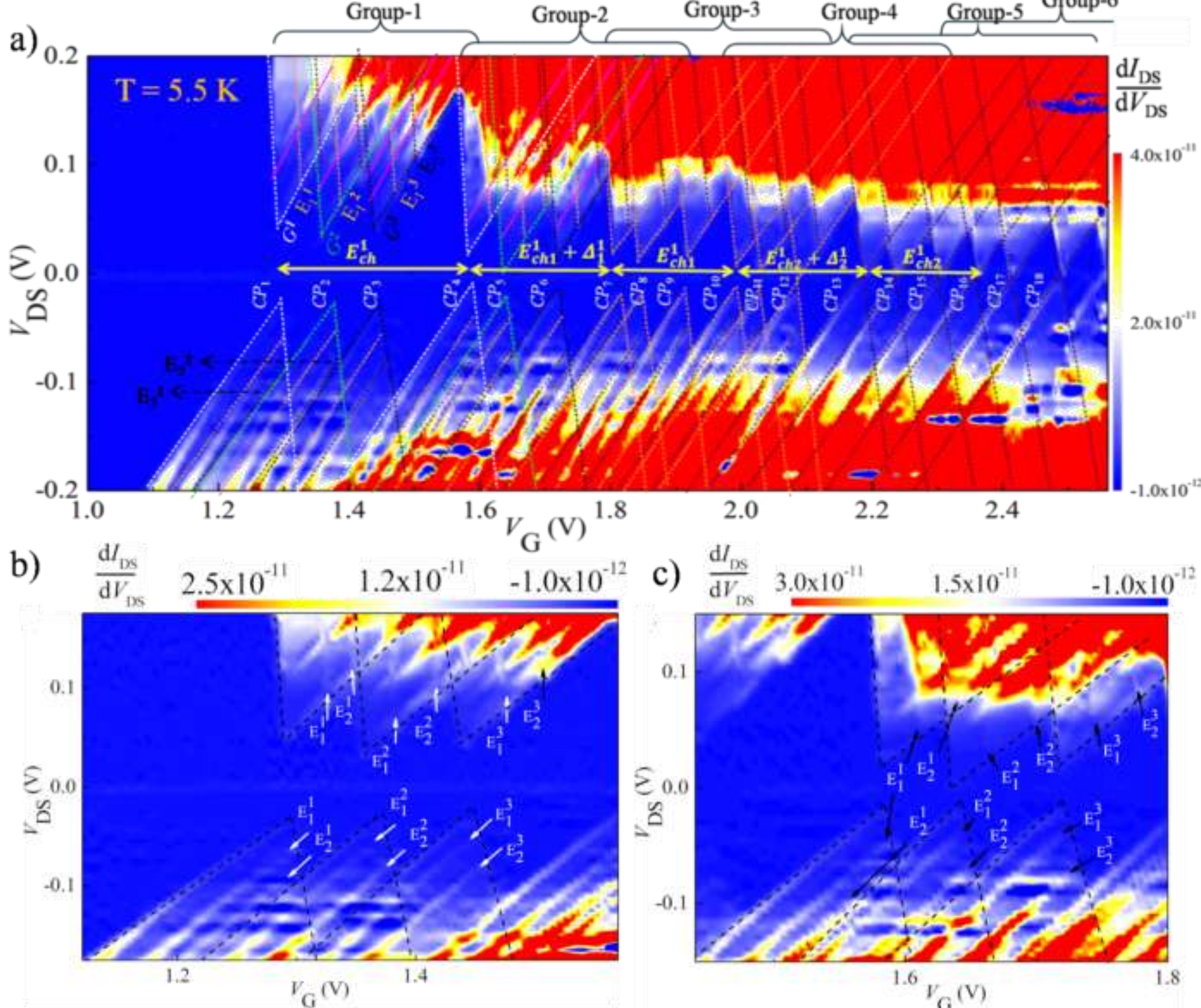

**Figure 2 Experimental electronic transport characteristics of the device. a,** The conductance is plotted in the $V_G$-$V_{DS}$ plane at $T$ = 5.5 K. The observed successive Coulomb oscillations within Group-1 correspond to single-electron transport through the ground states of three different donor molecules, molecule-1, molecule-2, and molecule-3, respectively, where the successive ground-state features are highlighted by white ($G^1$), green ($G^2$) and black ($G^3$) dotted lines, respectively. The dark-yellow ($E_1^1, E_1^2, E_1^3$ ) and brown ($E_2^1, E_2^2, E_2^3$) dotted lines running parallel to the left-edges of the diamonds are the signatures of simultaneous electron transport through 1st and 2nd excited states of each molecule. Next successive groups of Coulomb oscillations, denoted by Group-1 to Group-6, correspond to electronic transport through the higher charge states and excited states of the donor-molecules, where 1st, 2nd and 3rd Coulomb peaks of each group correspond to electron transport through the isolated molecule-1, molecule-2 and molecule-3, respectively. The systematic evolutions of the color schemes of the electronic states represent the corresponding evolution of the different donor-molecule energy levels with the variation of applied bias-voltages. The transport lines, marked by pink, are the signatures of source LDOS which are observed only for the positive bias owing to the inherently asymmetric tunnel barriers at the source and drain side. The cross-biasing effect is also prominent on the initial few groups of Coulomb diamonds. **b-c,** Zoom-in view of the Group-1 and Group-2, respectively, where the white arrows point different excited-state signatures of the three triple-donor molecules.

Multiple-donors, as like molecules, is the most feasible scenario as supported by the probabilistic calculation (Fig. 1c).[38-40] In addition, the observation of several groups of Coulomb-peaks cannot be explained based on Mott-Hubbard impurity band,[39] where only two prominent bands (lower and upper Hubbard band) would be observed.

Thus, the observed periodic and repetitive pattern of Coulomb-peaks plausibly originated due to active single-electronic transport from multiple individual multi-donor-induced QDs. Moreover, prominent signature of excited state lines, running parallel to the left-edges of the Coulomb-peaks, is observed.[41-42] Notably, every Coulomb peaks within group-1, 2 host two excited states each, which systematically decrease by one for every two successive Groups. The boundaries within each of the Coulomb-peaks that signify the onset of electronic-transport through ground, first and second excited states of the QDs are denoted by $G$, $E_1$, $E_2$, respectively (Fig. 2). Correspondingly, the synonymous $V_G$-$V_{DS}$ co-ordinates are tabulated in Table-1.

Further, the extracted Lever-arm factors ($\alpha = C_G/C_G + C_S + C_D$) for $CP_1 - CP_3$ are found to be approximately 0.95, 0.94 and 0.928, respectively. The calculated energy-differences between $G - E_1$, i.e. $\Delta_1$, and between $E_1 - E_2$, i.e. $\Delta_2$, are also different for $CP_1$ - $CP_3$ as observed from Fig. 2 and tabulated in Table-1. Thus, $CP_1$, $CP_2$, $CP_3$ are indeed from three different QDs, namely, $QD_1$, $QD_2$, $QD_3$ respectively. Consequently, we denote the $G$, $E_1$, $E_2$, $\Delta_1$, $\Delta_2$ of the three QDs by $G^i$, $E_1^i$, $E_2^i$, $\Delta_1^i$, *and* $\Delta_2^i$, respectively, with $i$ defining the QD index.

**Table 1 Comprehensive description of the onset of single-electron transport through different energy levels of the three donor-molecules.** The corresponding values in the $V_G$-$V_{DS}$ plane are extracted from the experimental stability diagram of Fig 2. Based on these values, the corresponding charging-energies and energy separations between ground and different excited states of the isolated molecules are calculated, based on the concept and formalism of QCB.

| | | Peak No. | 1 | 4 | 7 | 10 | 13 | 16 |
|---|---|---|---|---|---|---|---|---|
| **Molecule 1** | Approximate Peak Positions in mV for | $G^1\ (V_G, V_{DS})$ | 1290, 50 | 1580, 20 | 1800, 20 | 1988, 10 | 2188, 0 | 2362, 0 |
| | | $E_1^1\ (V_G, V_{DS})$ | 1328, 85 | 1618, 55 | 1830, 45 | 2022, 35 | Not Observed | Not Observed |
| | | $E_2^1\ (V_G, V_{DS})$ | 1356, 110 | 1646, 80 | Not Observed | Not Observed | Not Observed | Not Observed |
| | | $\Delta_1^1 = E_1^1 - G^1 = 35$ meV | $E_{ch}^1 = 280$ meV | | $E_{ch1}^1 = 155$ meV | | $E_{ch2}^1 = 120$ meV | |
| | | $\Delta_2^1 = E_2^1 - E_1^1 = 25$ meV | | $E_{ch1}^1 + \Delta_1^1 = 190$ meV | | $E_{ch2}^1 + \Delta_2^1 = 145$ meV | | |
| | | **Peak No.** | **2** | **5** | **8** | **11** | **14** | **17** |
| **Molecule 2** | Approximate Peak Positions in mV for | $G^2\ (V_G, V_{DS})$ | 1358, 40 | 1636, 5 | 1842, 15 | 2026, 0 | 2236, 0 | 2414, 0 |
| | | $E_1^2\ (V_G, V_{DS})$ | 1386, 65 | 1664, 30 | 1884, 50 | 2070, 35 | Not Observed | Not Observed |
| | | $E_2^2\ (V_G, V_{DS})$ | 1422, 100 | 1704, 65 | Not Observed | Not Observed | Not Observed | Not Observed |
| | | $\Delta_1^2 = E_1^2 - G^2 = 25$ meV | $E_{ch}^2 = 267$ meV | | $E_{ch1}^2 = 150$ meV | | $E_{ch2}^2 = 115$ meV | |
| | | $\Delta_2^2 = E_2^2 - E_1^2 = 35$ meV | | $E_{ch1}^2 + \Delta_1^2 = 175$ meV | | $E_{ch2}^2 + \Delta_2^2 = 150$ meV | | |
| | | **Peak No.** | **3** | **6** | **9** | **12** | **15** | **18** |
| **Molecule 3** | Approximate Peak Positions in mV for | $G^3\ (V_G, V_{DS})$ | 1438, 50 | 1720, 25 | 1916, 20 | 2084, 0 | 2302, 0 | 2488, 0 |
| | | $E_1^3\ (V_G, V_{DS})$ | 1466, 75 | 1750, 50 | 1954, 50 | 2124, 30 | Not Observed | Not Observed |
| | | $E_2^3\ (V_G, V_{DS})$ | 1500, 105 | 1784, 80 | Not Observed | Not Observed | Not Observed | Not Observed |
| | | $\Delta_1^3 = E_1^3 - G^3 = 25$ meV | $E_{ch}^3 = 260$ meV | | $E_{ch1}^3 = 145$ meV | | $E_{ch2}^3 = 115$ meV | |
| | | $\Delta_2^3 = E_2^3 - E_1^3 = 30$ meV | | $E_{ch1}^3 + \Delta_1^3 = 170$ meV | | $E_{ch2}^3 + \Delta_2^3 = 145$ meV | | |

Correlating the identical characteristics, along-with striking similarities, in $\Delta_1^1$ and $\Delta_2^1$ values between $CP_1$ and $CP_4$, they must be the first and second charge states, respectively, of the ground-state energy level for $QD_1$. The corresponding charging-energy, extracted from the intersections of the extended slopes of $CP_1$ and $CP_4$, is found to be ~280 meV (Figs. 2; Table1) and the corresponding ground state Bohr radius is $r_B \approx 0.43$ nm; calculated utilizing the relation $E_{ch} = e/C_\Sigma$. $C_\Sigma$ is the self-capacitance of the dot given by $C_\Sigma = 4\pi\varepsilon_{Si} r_B$ with $\varepsilon_{Si}$ being the permittivity of the Si ≈ 11.9.[22] Moving along the $V_G$-axis, $\Delta_1^1$ for $CP_7$ and $CP_{10}$ exactly resemble $\Delta_2^1$ of $CP_1$, $CP_4$, affirming them to be 3rd and 4th charge state, respectively, of $QD_1$. This is also consistent with the presence of a-single excited state in $CP_7$ and $CP_{10}$, in contrast to the two excited states observed in $CP_1$ and $CP_4$. The respective addition energies would thus precisely be $E_{ch1}^1 + \Delta_1^1$ and $E_{ch1}^1$ (Fig. 2 and Table 1), signifying clear signatures of QCB. The corresponding Bohr radius is calculated to be $r_B \approx 0.77$ nm. Further correlating the systematically decremental charging energy, $CP_{13}$ and $CP_{16}$ would be the 5th and 6th charge states, consistent with the absence of any excited state signatures within, of the $QD_1$ with respective addition energies of $E_{ch2}^1 + \Delta_2^1$ and $E_{ch2}^1$. Corresponding $r_B \approx 1.01$ nm.

Similarly, $CP_2$ and $CP_5$ arise due to single-electron transfer through the 1st and 2nd charge states of $QD_2$ with charging-energy $E_{ch}^2$ and $r_B \approx 0.452$ nm. $CP_3$ and $CP_6$ correspond to the 1st and 2nd charge states of $QD_3$ with charging-energy $E_{ch}^3$ and $r_B \approx 0.464$ nm. $CP_8$ and $CP_{11}$ corresponds to the 3rd and 4th charge state configurations of $QD_2$, with addition energies $E_{ch1}^2 + \Delta_1^2$, $E_{ch1}^2$, respectively and $r_B \approx 0.805$ nm. On-the-same-note, $CP_9$ and $CP_{12}$ represent the 3rd and 4th charge state of $QD_3$, with addition energies $E_{ch1}^3 + \Delta_1^3$, $E_{ch1}^3$, respectively and $r_B \approx 0.833$ nm. Consistent with the dynamics, $CP_{14}$ and $CP_{17}$ are the 5th and 6th charge states of the $QD_2$ with respective addition energies of $E_{ch2}^2 + \Delta_2^2$ and $E_{ch2}^2$, whereas, $CP_{15}$ and $CP_{18}$ are the 5th and 6th charge states of the $QD_3$ with addition energies of $E_{ch2}^3 + \Delta_2^3$ and $E_{ch2}^3$, respectively. The corresponding $r_B$ for both QD2 and QD3 in these charge states are ≈1.05 nm.

The observed small $r_B$ of the QDs, coupled with significant energy level separations and high charging-energies

(Table-1), strongly indicate that they are plausibly originated from multi-donor interactions. Presence of three distinctly separated energy levels within each QD suggests that three closely spaced P-donors interacted to form each of the QDs. Additionally, the absence of any modulations of the Coulomb peak edges while crossing each-other, demands the QDs are non-interacting.[8]

The observed systematic decrement of charging-energies for the successive higher energy levels of the QDs are reasonable as the electronic wave-functions become more spatially delocalized, leading to reduced effective Coulomb repulsion. This delocalization, coupled with the larger spatial extent of higher energy orbitals, increases the effective volume over which the electron charge is distributed.[43] Thus, coupling of the QDs with the leads and correspondingly the capacitances increase[44-45] leading to decreased charging energies followed by a similar trend in the observed $\alpha$-values for successive Coulomb-peaks, which reaches approximately 0.67, 0.646 and 0.62, for $CP_{16}$-$CP_{18}$ respectively. This could also be attributed to the cross bias effect[32] and larger coupling of higher energy-levels of the QD to the leads.[44] So, the calculated charging-energies of the QD ground-states and the 1st-excited states may be over-estimated.

Using these α-values, the estimated positions of the three non-interacting QDs, from the source reservoir, are ≈111 nm, ≈108 nm, ≈118 nm, respectively[35] and the corresponding ionization energies are be ≈840 meV, ≈780 meV and ≈700 meV considering the conduction band edge to be ≈2560 mV.

Additional faint lines are observed within the first two Coulomb-peak groups, running parallel to their right edges (pink-lines in Fig. 2).These features are visible only for the positive bias condition, and can be ascribed to random fluctuations in local density of states (LDOS) of the source reservoir.[46-48] The absence of these lines in the negative bias can be attributed to the intentionally fabricated asymmetric tunnel barriers formation.[49]

Moreover, to study the high-temperature response of the device, we have discussed in detail the temperature-dependent variation of $|I_{DS}|$ against gate-voltage ($V_G$) in the supplementary-information (Fig. S1a). The isolated Coulomb-peaks, especially at the lower $V_G$ region, clearly sustain at-least upto $T$ = 60 K. Further, the coalescence of the Coulomb-peaks at higher temperatures is aggravated due to the overlapping of individual single-electronic-peaks arising from the different triple-donor molecules. Otherwise, if only a single triple-donor molecule would have been present within the device nano-channel, the corresponding isolated Coulomb-peaks should have been sustained even upto $T$ = 300 K, which is explained in detail within the supplementary-information through subsequent simulations (Fig. S3).

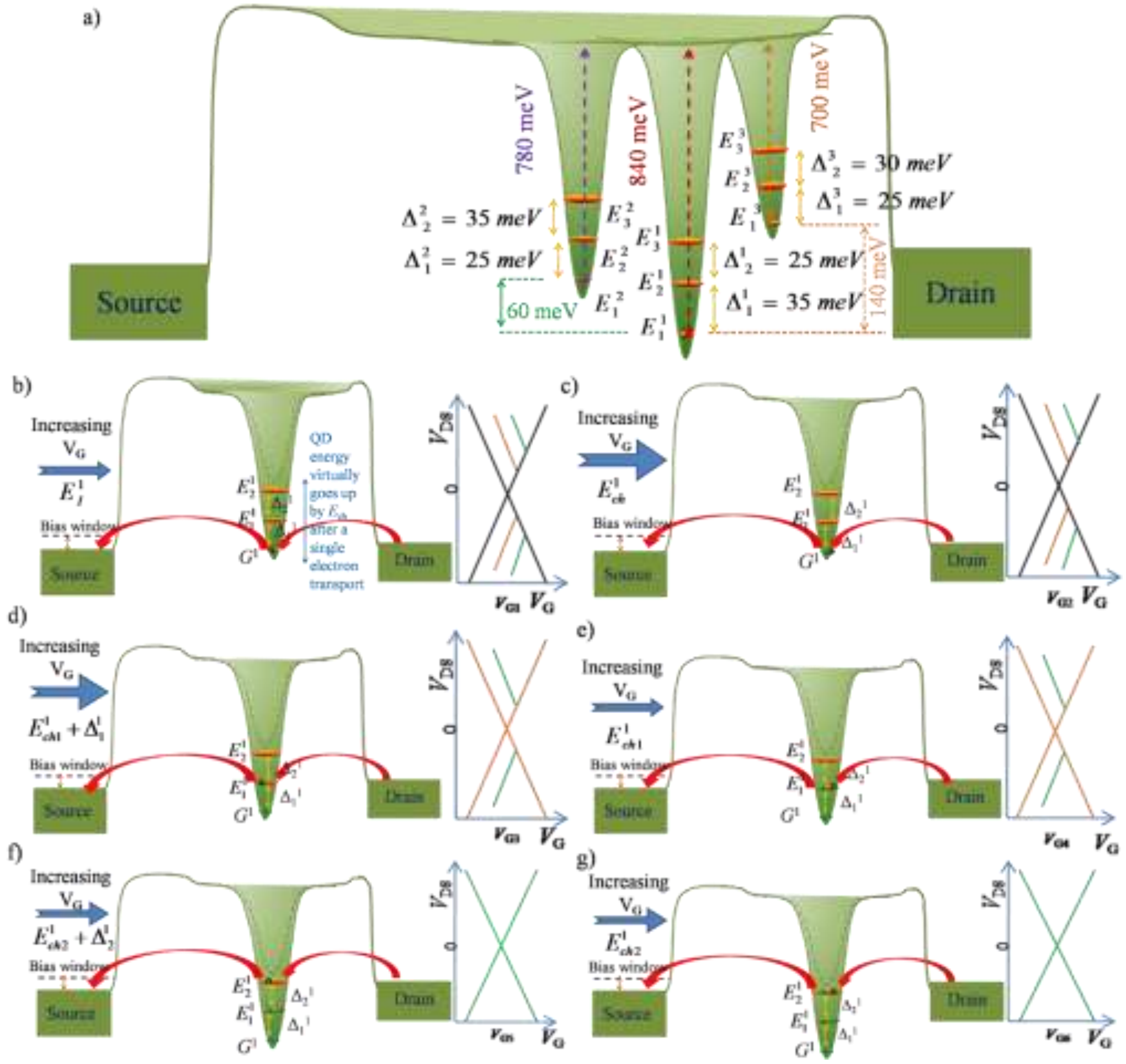

**Figure 3 Schematic representation of QCB transport. a,** Deduced energy-level scheme of the three isolated-donor molecules. **b-g,** One-by-one electron transport and filling in the different charge states of successive energy levels within one of the three isolated transport-active donor-molecules, namely molecule-1, with the evolution of gate bias voltage. For the electron-transfer schematic, we have considered the drain terminal is negatively biased. So, the drain Fermi-level is higher than the grounded source-terminal and the single electrons tunnel from drain to source. The expected particular Coulomb peak, with the signature of the participating excited states in the electronic transport, associated with the transport via each of the energy levels is also depicted.

## C. QCB mediated molecular-filling of single electrons

To decipher the detailed QCB transport mechanism, we describe the electron filling process within the observed potential landscape, as presented schematically in Figs. 3a-g with the experimentally measured energy scales. We analyze the sequential occupation of discrete electronic states of molecule-1, one of the three transport-active molecules.

The electronic transport initiates when the lowest-lying and empty $G^1$ state aligns with the bias window through gate voltage adjustment (Fig. 3b). Consequently, the peak $CP_1$ arises (Fig. 3b) and is due to the 1st charge state of $G^1$. Further increasing $V_G$, $G^1$ surpasses the source Fermi level and a single electron is trapped in the $G^1$ ; consequently, the energy of molecule-1 virtually enhances by $E_{ch}^1$ (marked by black dotted line in Fig. 3c) corresponding to the ground state, leading to the transport being blockaded. Applying $V_G$ equivalent to $E_{ch}^1$, the singly-filled $G^1$ level, in the 2nd charge state configuration, realigns within the bias window (Fig. 3c) permitting another single-electron transport and $CP_4$ is observed; the Pauli Exclusion Principle prevents further electron transport via this ground state. Further electronic transport through the doubly-occupied molecule-1 can be facilitated by aligning the next empty $E_1^1$ level, $\Delta_1^1$ away from $G^1$ with the bias window which would require corresponding charging-energy ($E_{ch1}^1$) added with $\Delta_1^1$ to be provided and the Coulomb-peak $CP_7$ would be observed (Fig. 3d).To lift the next blockade, a further increment of the molecule-1 potential by $E_{ch1}^1$ is required and,electron will tunnel via the 2nd charge state of $E_1^1$ for the triply-occupiedmolecule-1 (Fig. 3e) to facilitate $CP_{10}$. Similarly, the next unoccupied energy level $E_2^1$ would actively participate in the transport with the 1st and 2nd charge state configurations upon successively applying gate voltages equivalent to energy $E_{ch2}^1 + \Delta_2^1$ and $E_{ch2}^1$ and we will observe $CP_{13}$, $CP_{16}$ respectively. This is schematically presented in Figs. 3f-g. Identically, the other two molecules, molecule-2 and molecule-3, facilitate systematic electron transport based on the concept of QCB and the overall experimental transport characteristics of Fig. 2 is observed.

## D. First-principles insights into molecular-orbital structures

To obtain a qualitative understanding of the observed energy spectra and the nature of the energy levels participating in the transport, we perform first-principles Density Functional Theory simulations of silicon nano-structures containing three closely spaced P-donors using Quantum-ATK software.[30], as shown in Fig. 4a.

Figure 4b demonstrates the total-density-of-states (TDOS) spectra of the 3-donors, individually and cumilitively, where the lowest three energy levels, marked by $G$, $E_1$, $E_2$ respectively, qualitatively correspond to the experimentally observed lowest-lying energy levels of either of the three molecules that are actively participating in the transport.
This splitting is consistent with the expectation that the number of split levels within the GS manifold correlates with the number of strongly coupled donors. The separation between these ground-state levels qualitatively matches experimental observations and have almost the same ratio of $\Delta_1^2 : \Delta_2^2 : E_I^2 \equiv 1 :1.4:33$, reasonably close to the spectra for molecule-1 and molecule-2. Therefore, this is the plausible configuration of the three individual donors that constitutes the donor-molecules. The projected-density-of-states (PDOS) at the location of each P donor are also shown in Fig. 4c, where different colors indicate contributions of *s*, *p*, *d* orbitals of the left, middle and right donors at their respective locations. Fig. 4d signifies the relative weight of donor states against the states of the whole system[50-51] and corresponding ionisation energy of the donor-molecule is found to be ≈ 1.2 eV. Furthermore, the orbital-decomposed PDOS of the donors indicates that the split ground-state levels are primarily-derived from the *s* orbitals of the donors whereas, the third level of ground-state-manifold and further excited-states of the simulated-spectra are resulting from hybridization of the *s*, *p*, and *d* orbitals which is qualitatively different from simple muti-donor-clusters where, the ground-state-manyfolds originate mostly due to interactions from s-orbitals only.[25]

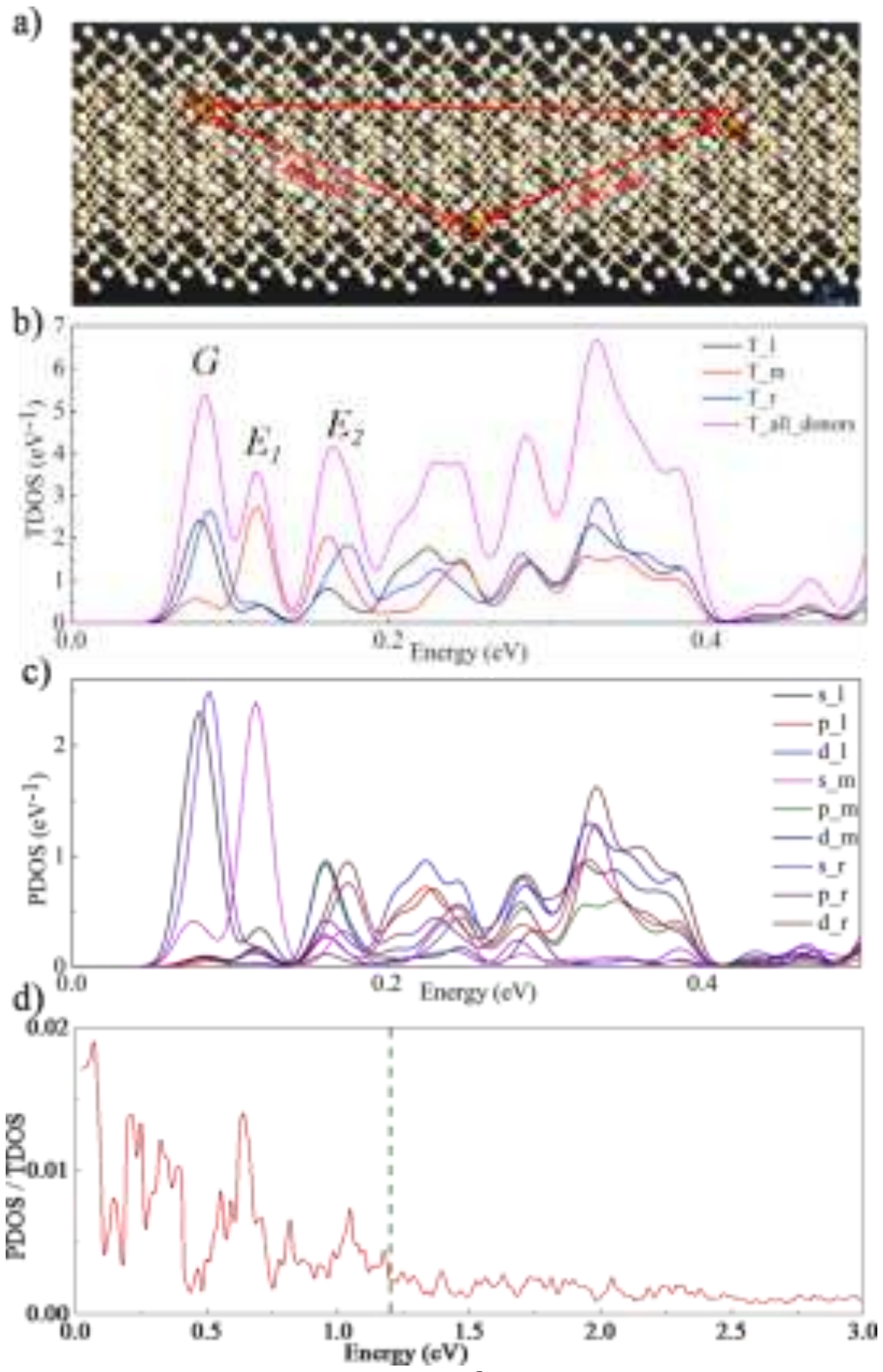

**Figure 4 First-principles analysis of nanostructures containing 3-P donors.**The DFT calculations were carried out using the Linear Combination of Atomic Orbitals (LCAO) method, with the Generalized Gradient Approximation (GGA) approach to handle exchange-correlation interactions, employing the PBE Double Zeta Polarized functional. The k-point density was set to (1 x 1 x 4), with a grid mesh cutoff of 45 Hartree. Geometry optimization was performed using the LBFGS optimizer, with a self-consistent field iteration tolerance of 0.0001 Hartree. During optimization, lattice parameters and atomic positions were adjusted until atomic forces were reduced to below 0.05 eV/Å. **a,** Atomistic view of of the simulated nanostructure, containing P-donors (distance between $P_1$–$P_2$ and $P_2$–$P_3$ is ~2.06 nm and ~1.89 nm, respectively). **b,** The corresponding TDOS spectrum contributions from each donor of the molecule and the overall TDOS spectrum. *l*, *m*, *r* defines the left, middle, right donor atoms respectively. $\Delta_1$, $\Delta_2$ are found to be ≈34 meV and ≈48 meV**,** respectively. **c,** Respective individual orbital contributions from each donor, constituating the TDOS. *s*, *p*, *d* suggests the atomic orbitals. **d,** PDOS / TDOS spectra of the P-donor-molecule. The ionisation energy is estimated by taking the ratio of PDOS/TDOS, which signifies the relative weight of donor states against the states of the whole system. In the lower energy range, the contribution from the donor is dominated but, the contribution is trivial at higher energy range (above the green dashed line) due to the inclusion of more number of silicon atoms. The ionisation energy is calculated from the point where this ratio saturates close to 0 and found to be around 1.2 eV.

## E. Simulation for the observed QCB effects

To investigate the validity of the actual scenario within the theoretical framework, we construct an equivalent-capacitor model with three isolated donor-molecules involved in electron tunneling and reproduce the experimental stability diagram at $T$ = 5.5 K using modified rate equation approach[52] equipped for QCB (Figs. 5a–b). Within the simulated stability diagram (Fig. 5b), we observe traces of all the discrete energy levels, that actively participated in the experimental electron transport through the device, along with the LDOS features. Moreover, the systematic modulation of the charging-energies corresponding to the different energy levels of the three distinct molecules, are also reflected within the simulated results. The quantitative and qualitative similarities between the experimental and simulated stability diagrams (Fig. 2, Fig. 5b) and different energy scales like ionization and charging-energies confirm the validity of QCB single-electron transport through three-isolated donor-molecules. The slight qualitative mismatch in the initial few Coulomb peak shapes is most-likely originated due to the exclusion of cross-biasing effects[28] within the simulation. This does not, however, impact the overall conclusions.

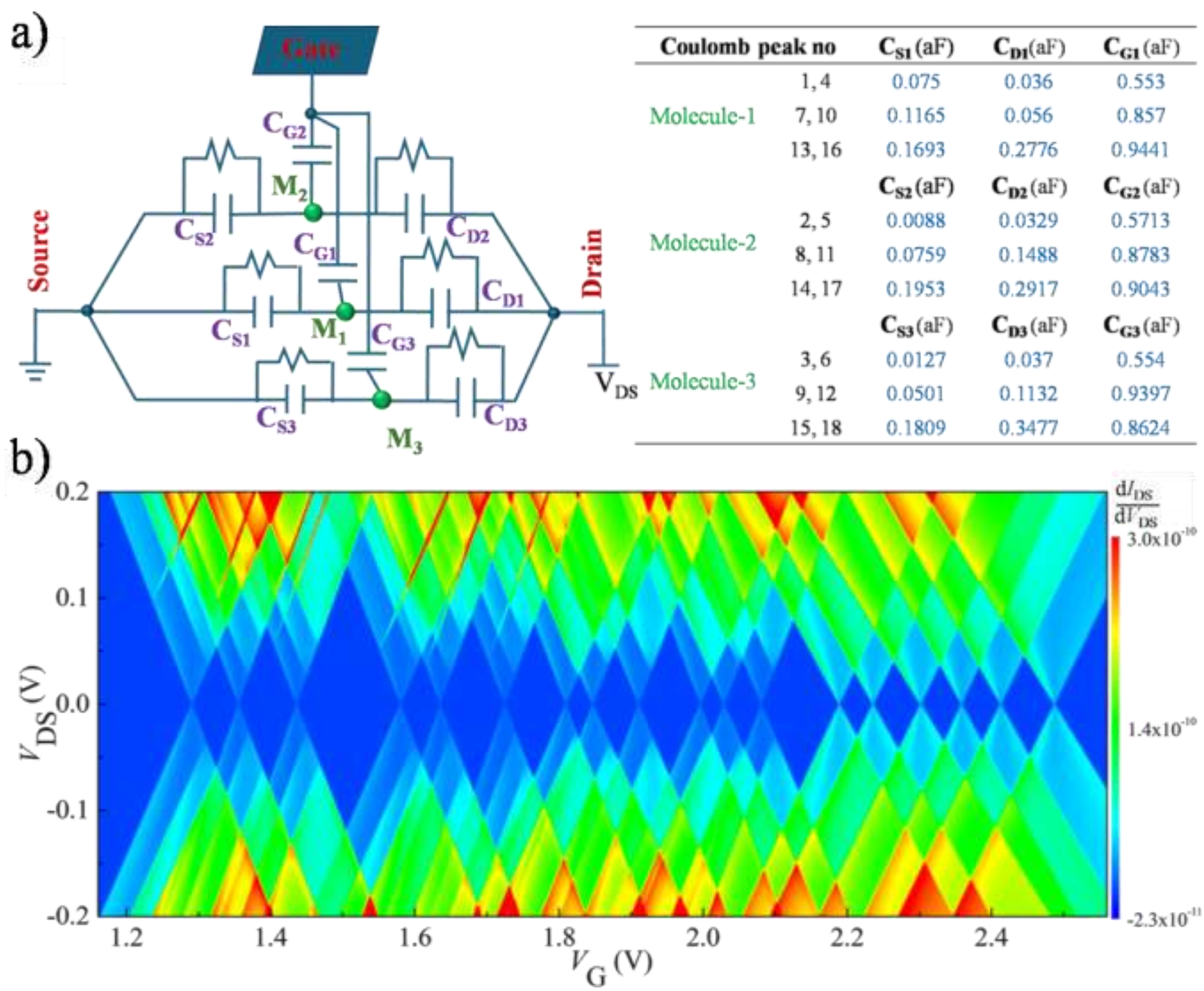

| | Coulomb peak no | $C_{S1}$ (aF) | $C_{D1}$ (aF) | $C_{G1}$ (aF) |
|---|---|---|---|---|
| | 1, 4 | 0.075 | 0.036 | 0.553 |
| Molecule-1 | 7, 10 | 0.1165 | 0.056 | 0.857 |
| | 13, 16 | 0.1693 | 0.2776 | 0.9441 |
| | | $C_{S2}$ (aF) | $C_{D2}$ (aF) | $C_{G2}$ (aF) |
| | 2, 5 | 0.0088 | 0.0329 | 0.5713 |
| Molecule-2 | 8, 11 | 0.0759 | 0.1488 | 0.8783 |
| | 14, 17 | 0.1953 | 0.2917 | 0.9043 |
| | | $C_{S3}$ (aF) | $C_{D3}$ (aF) | $C_{G3}$ (aF) |
| | 3, 6 | 0.0127 | 0.037 | 0.554 |
| Molecule-3 | 9, 12 | 0.0501 | 0.1132 | 0.9397 |
| | 15, 18 | 0.1809 | 0.3477 | 0.8624 |

**Figure 5 Simulation of the overall transport characteristics within modified rate equation approach. a,** Schematic equivalent capacitor circuit diagram of the studied device. Blue dots represent nodes in the circuit connection whereas the green spheres represent the three isolated donor molecules that actively participate in the single-electron transport. **b,** The corresponding simulated stability diagram at $T$ = 5.5 K for the same configuration.

## IV. Conclusions

The experimentally demonstrated systematic filling of several electrons into the orbital-resolved molecular levels synchronizes with the observed clear and sustained quantum Coulomb blockade effect correlated with the systematic evolution of the distinguished molecular excited-states. Correspondingly, the observed systematic reduction of charging-energies with higher orbital occupation is the manifestation of increased orbital Bohr radii and charge delocalization. Complemented by donor-potential simulations, first-principle DFT calculations and quantitative rate-equation simulations of the experimental stability diagram, we have demonstrated the experimental realization and comprehensive characterization of triple-phosphorous-donor molecule's electronic spectrum and intrinsic quantum behavior. Further citing the experimentally observed charging-energies of few hundred meV and ionization potential of around 700 meV, we have demonstrated the plausible room temperature operation of such donor-molecule based quantum devices. These are supported by the experimental temperature dependent gate-voltage characteristics and subsequent rate equation simulations of corresponding high-temperature stability diagrams. Albeit, the relatively smaller separations between the higher energy levels may limit the coherence at elevated temperatures.

## Data availability

The data that support the findings of this study are available within the article.

## Acknowledgement

This work was supported by the project STARS-2/2023-0715 funded by IISc-MHRD, India, and SR/FST/PS-II/2019/84 funded by DST, India. P.S. and S.C. acknowledge the CSIR and Ministry of Education, India, respectively, for the research fellowship.

# Quantum Coulomb Blockade in Orbital Resolved Phosphorus Triple-Donor Molecule

Soumya Chakraborty,*[1] Pooja Sudha,*[1] Hemant Arora,[1] Daniel Moraru,[2] and Arup Samanta[1,3,a)]

[1]Quantum / Nano-science and Technology Lab, Department of Physics, Indian Institute of Technology Roorkee, Roorkee 247667, Uttarakhand, India
[2]Research Institute of Electronics, Shizuoka University, 3-5-1 Johoku, Chuo-ku, Hamamatsu 432-8011, Japan
[3]Centre of Nanotechnology, Indian Institute of Technology Roorkee, Roorkee 247667, Uttarakhand, India
[a)]arup.samanta@ph.iitr.ac.in
* Equal Contribution

## Temperature dependent gate voltage characteristics:

To understand the feasibility of high temperature (upto $T$ = 300 K) single-electronic operation through such a device under consideration, we have measured the gate voltage response of the source-drain current at different temperatures and is shown in Fig. S1a.

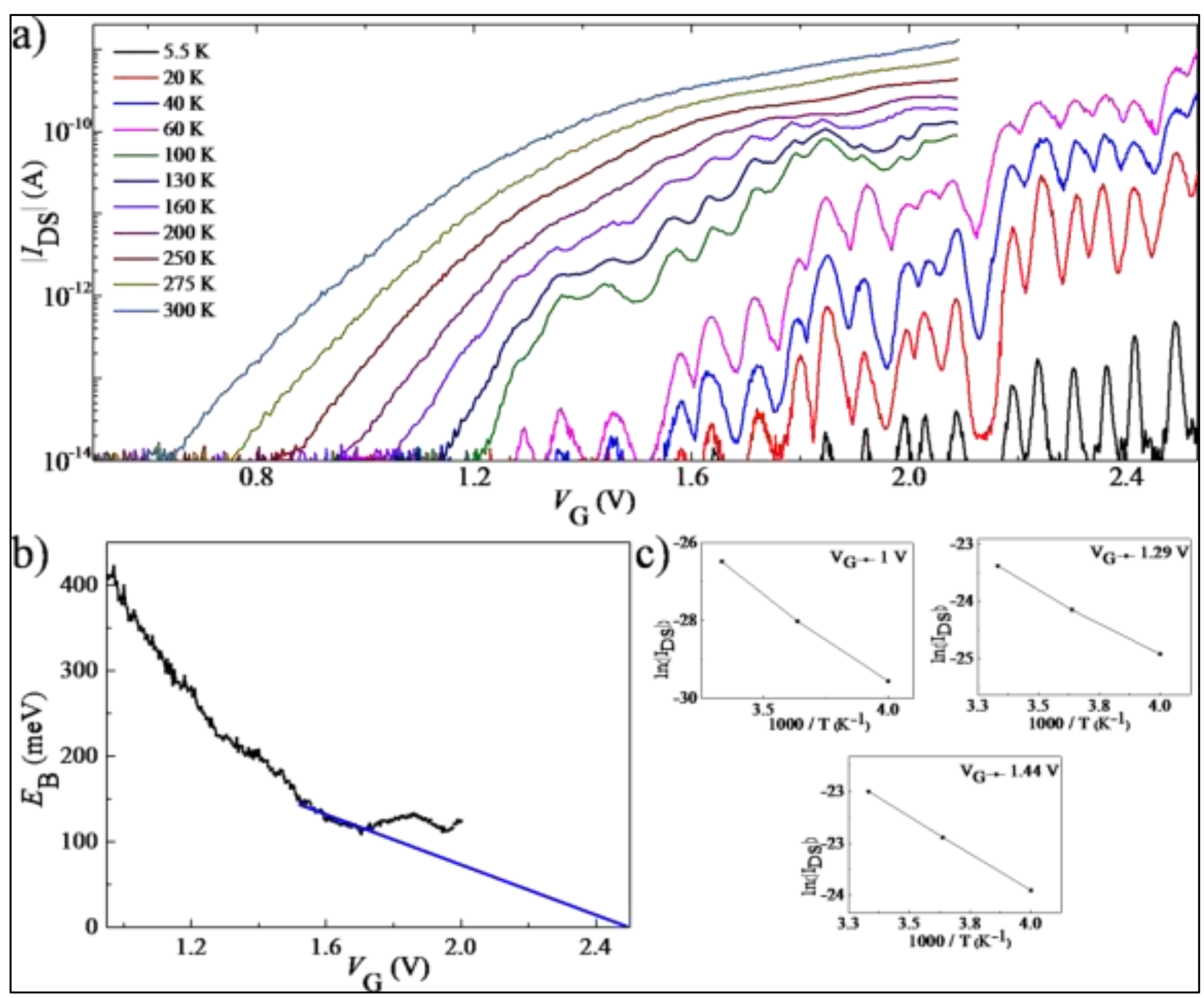

**Figure S1a,** Temperature dependent gate voltage characteristics of the device. **b,** Modulation of the effective barrier heights with the variation in gate voltage extracted by exploiting the $|I_{DS}|$ - $V_G$ characteristics, at $T$ = 250 K, 275 K and 300 K. **c,** Arrhenius plot at some selected $V_G$ values.

We can clearly observe the systematic evolution of the Coulomb peaks with increasing temperature. The initial few peaks that appeared in the experimental stability diagram (Fig. 2 of the main text) at higher $V_{DS}$ values gradually appear at low $V_{DS}$ of 5 mV at higher temperature which is expected.[1-2] The clear Coulomb oscillations sustain at higher temperatures, upto the range of $T$ = 200K. But, the fine features such that the excited state signatures are not clearly identifiable even at temperatures around $T$ = 60 K. This plausibly be ascribed to the thermal broadening, ≈ 3.5 $k_BT$ (approximately 20 meV at $T$ = 60 K; comparable to the spacing between discrete energy levels of the multi-donor molecules),[3-4] which may smears off the discreteness among the energy levels at elevated temperatures. But there should be some other factors like the merging of several individual current peaks originating from the three different multi-donor clusters, described in detail later, that should also greatly contribute to the dilution of the excited state features at such temperature range. Moreover, some very-faint oscillations can be observed even at temperatures $T$ = 275 K and 300 K around $V_G$ = 1.6 V, 1.9 V. It clearly suggests that, even at such high temperatures, the electronic transport through the device is not yet completely thermally activated over the barrier electron propagation.[5] So, if we extract the effective barrier height (Fig. S1b) from the temperature-dependent gate voltage characteristics exploiting Arrhenius plot (Fig. S1c), that would be under-estimation. This is reflected in the extracted barrier heights, as plotted in Fig. S1b; which are substantially lower than the calculated ionization energies (≈ 700 meV – 800 meV) for the individual multi-donor molecules, as mentioned in the main text.

**Scope of the high temperature single electronic operation:**

The temperature dependent $|I_{DS}|$ - $V_G$ characteristics illustrates significant smearing off for the peak to valley ratio of the individual single-electron peaks at elevated temperatures which is mainly due to the overlap of different current peaks originating from the three different multi-donor molecules that actively participate in the overall electrical transport through the device under consideration. This is further supported by the simulated Coulomb blockade at temperature $T$ = 80K, as shown in Fig. S2.

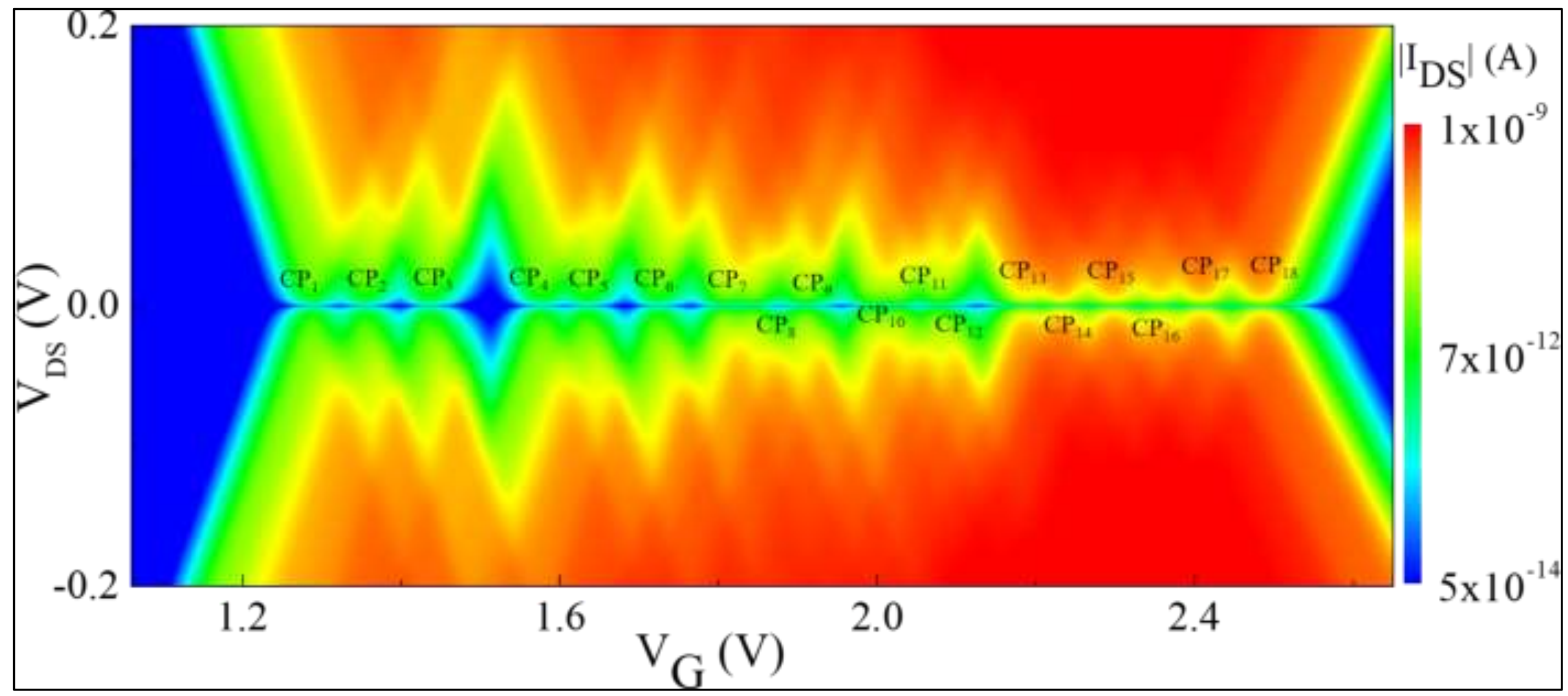

**Figure S2 Simulated stability diagram for the device under configuration at $T$ = 80 K.** The capacitance values used for the simulation are similar to the values mentioned in Fig 5a of the main text.

The simulation clearly depicts the effect of thermal broadening at elevated temperatures and overlapping of individual current peaks due to the simultaneous presence of three active, individual molecules. Due to these effects, the fine features like the excited states present within the Coulomb peaks (clearly observed in Fig. 2 and Fig. 5 of the main text), as well as the individuality of the isolated Coulomb peaks are not clearly visible in the simulated stability diagram despite their very large ionization and charging energies.

Instead of three individual three-donor molecules, if there would have been only a single three-donor molecule present within the selectively doped region then the corresponding Coulomb peaks should have had sufficient separation among them; at least at $T$ = 80 K, considering the thermal broadening and the large ionization and charging energies of the molecule, as mentioned in the main text. Moreover, the excited state features should have also been observed unlike in the experimental gate voltage characteristics shown in Fig. S1a. This is further asserted through the simulated Coulomb blockade at $T$ = 80 K, shown in Fig. S3a, considering only the presence of molecule 1within the system. In that context, such a multi-donor system would have been a great platform for qubit operation and other single electronic applications at least upto for 80 K technology.

If we further increase the temperature, the signatures of the excited states within the Coulomb peaks should have vanished beyond $T$ = 100 K due to the thermal broadening effect but the isolated nature of the Coulomb peaks, at least of the initial few, should have been sustained, even upto $T$ = 300 K considering the very high charging and ionization energies (mentioned in the main text). It is also supported by the simulated stability diagram at $T$ = 200 K, 300 K, considering only the presence of molecule 1within the system instead of three isolated three-donor molecules and is presented in Fig. S3b, c. So, multi-donor molecule based quantum devices could be a great platform for potential room temperature quantum applications like quantum sensing,[6-7] quantum metrology,[8] quantum memories,[9] single-photon detectors[10] etc.

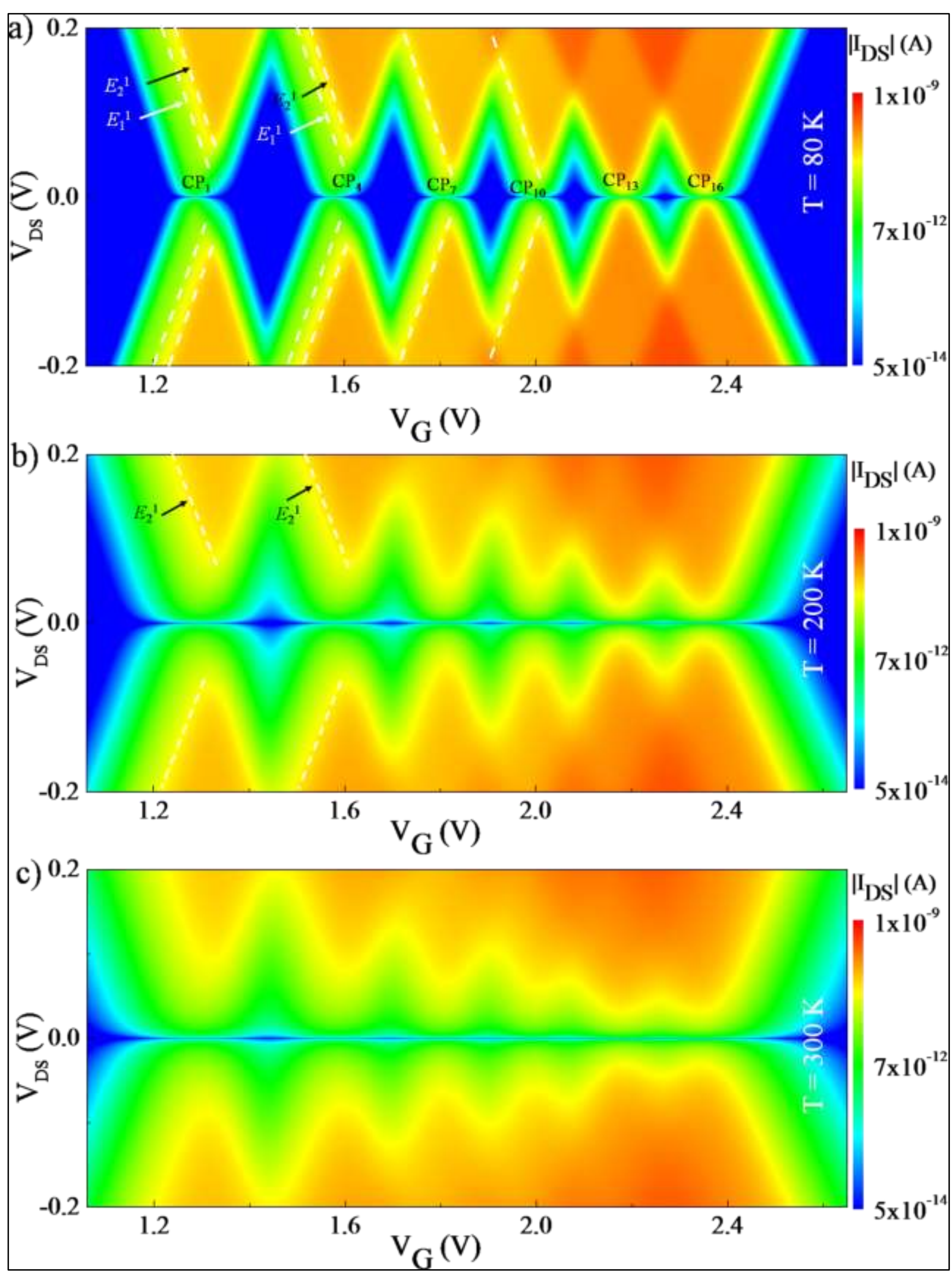

**Figure S3 Simulated stability diagram considering the thermal broadening when only molecule 1 is present in the selectively doped channel.** **a,** $T$ = 80 K **b,** $T$ = 200 K, **c,** $T$ = 300 K. Systematic dilution of the excited state visibility with increasing temperature is also prominent with increasing temperature. But the initial couple of overall Coulomb peaks are clearly visible with sustained Coulomb gap between them.